\documentclass[english,letterpaper,twocolumn,pra,aps]{revtex4}
\usepackage{graphicx}
\usepackage{amssymb}

\makeatletter

\large  

\input epsf

\makeatother

\usepackage{babel}
\makeatother
\begin{document}

\title{Aharonov-Bohm magnetism and Landau diamagnetism in semimetals}

\author{Eugene B. Kolomeisky$^{1}$\footnote{Author to whom any correspondence should be addressed.  Electronic address:  ek6n@virginia.edu} and Joseph P. Straley$^{2}$}

\affiliation
{$^{1}$Department of Physics, University of Virginia, P. O. Box 400714,
Charlottesville, Virginia 22904-4714, USA\\
$^{2}$Department of Physics and Astronomy, University of Kentucky,
Lexington, Kentucky 40506-0055, USA}

\begin{abstract}
We compute the magnetic response of hollow semimetal cylinders and rings to the presence of an axial Aharonov-Bohm magnetic flux, in the absence of interactions.  We predict nullification of the Aharonov-Bohm effect for a class of dispersion laws that includes "non-relativistic" dispersion and demonstrate that at zero flux the ground-state of a very short "armchair" graphene tube will exhibit a ferromagnetic broken symmetry.  We also compute the diamagnetic response of bulk semimetals to the presence of a uniform magnetic field, specifically predicting that the susceptibility has a logarithmic dependence on the size of the sample.   
\end{abstract}


\maketitle

\section{Introduction}

Ajiki and Ando (AA) \cite{Ando} and Kane and Mele \cite{Kane} have observed that the long-wavelength low-energy dynamics of the electrons of graphene \cite{RMP} when confined to a cylindrical surface is described by the two-dimensional massless Dirac equation, and that the effects of the tube size and its chirality can be represented by a fictitious vector potential.  If additionally there is an axial Aharonov-Bohm (AB) flux $\Phi$ present, then the energy eigenvalues of the Dirac equation are given by  \cite{Ando}
\begin{equation}
\label{eigenvalues}
E_{n}(q_{z})=\pm \gamma\left [q_{z}^{2}+\left (\frac{2\pi}{W}\right )^{2}(n+\phi\pm\alpha)^{2}\right ]^{\frac{1}{2}}
\end{equation}
where the overall upper and lower signs refer to the conduction and valence bands (respectively), $\gamma$ is the Fermi velocity $v_{F}$ times $\hbar$, $q_{z}$ is the wave vector in the axial direction, $W$ is the cylinder circumference, $n=0, \pm1, ...$ is the azimuthal quantum number, and $\phi=\Phi/\Phi_{0}$ is the dimensionless AB flux measured in units of the flux quantum $\Phi_{0}=hc/e$.  The fictitious flux parameter $\alpha$ combines the effects of winding and curvature, and provides a classification of nanotubes \cite{Ando,Kane}.  It has opposite signs in the $K$ and  $K'$ Dirac valleys \cite{Ando} as indicated by the lower and upper signs in front of $\alpha$ in Eq.(\ref{eigenvalues}).  This flux is not due to a physical magnetic field and for $\phi=0$ the system  is time-reversal symmetric.   All nanotubes can be classified as being semimetallic "armchair" ($\alpha=0$), insulating ($\alpha$ close to $1/3$), or semiconducting ($\alpha\ll1$) \cite{Ando,Kane}.

Since the integer part of $\alpha$ or $\phi$ can be absorbed into the definition of the azimuthal quantum number $n$, any physical property is a periodic function of $\alpha$ or $\phi$ with unit period.  As an example of such a property, AA \cite{Ando} calculated the AB magnetic response of an undoped cylinder, which is a conceptually interesting problem because at zero temperature the free carriers are absent and the effect is entirely due to the electrons of the filled Dirac sea.  Additionally, graphene, while being a semimetal, represents a marginal case between normal metals (where AB magnetism is experimentally interpretable in terms of persistent currents \cite{Kulik}) and insulators (where the effect is expected to be suppressed due to the band gap).  For a non-chiral $\alpha=0$ tube of length $L$, the valence electrons of given spin and belonging to the $K$-valley contribute to the ground-state energy the quantity 
\begin{equation}
\label{E_K}
\mathcal{E}_{K}^{\textrm{\textrm{(cyl)}}}(\phi)=-\gamma \sum_{n=-\infty}^{\infty}\int_{-\infty}^{\infty}\frac{Ldq_{z}}{2\pi} \left [q_{z}^{2}+\left (\frac{2\pi}{W}\right )^{2}(n+\phi)^{2}\right ]^{\frac{1}{2}}
\end{equation}
The corresponding magnetic moment $\mathcal{M}_{K}^{\textrm{\textrm{(cyl)}}}(\phi)$ and differential susceptibility $\chi_{K}^{\textrm{\textrm{(cyl)}}}(\phi)$ are determined by differentiation of Eq.(\ref{E_K}):
\begin{equation}
\label{M_K_chi_K}
\mathcal{M}_{K}^{\textrm{\textrm{(cyl)}}}(\phi)=-\frac{W^{2}}{4\pi\Phi_{0}} \frac{\partial E_{K}^{\textrm{\textrm{(cyl)}}}}{\partial \phi},\chi_{K}^{\textrm{\textrm{(cyl)}}}(\phi)=\frac{W^{2}}{4\pi\Phi_{0}}\frac{\partial \mathcal{M}_{K}^{\textrm{\textrm{(cyl)}}}}{\partial \phi}
\end{equation}
For a generic tube the ground-state energy (accounting for the spin degeneracy and including the contributions of both valleys) can be written in terms of $\mathcal{E}_{K}^{\textrm{\textrm{(cyl)}}}$ as
\begin{equation}
\label{E_general}
\mathcal{E}^{\textrm{\textrm{(cyl)}}}(\phi,\alpha)=2(\mathcal{E}_{K}^{\textrm{\textrm{(cyl)}}}(\phi+\alpha)+\mathcal{E}_{K}^{\textrm{\textrm{(cyl)}}}(\phi-\alpha)), 
\end{equation}
and similar relationships hold for the total magnetic moment and susceptibility, with $\mathcal{E}$ replaced by $\mathcal{M}$ or $\chi$.  Thus by computing only one of the $K$-functions as a function of the flux $\phi$ we can understand the general problem.  

In what follows we will be also interested in the one-dimensional \textrm{\textrm{(ring)}} version of the same problem that is obtained by setting $q_{z}\equiv0$ in the spectrum Eq.(\ref{eigenvalues}).  Then instead of Eq.(\ref{E_K}) the ground-state energy is given by
 \begin{equation}
\label{E_K_ring}
\mathcal{E}_{K}^{\textrm{\textrm{(ring)}}}(\phi)=-\gamma \sum_{n=-\infty}^{\infty} \frac{2\pi}{W}|n+\phi|
\end{equation} 
and relationships analogous to those of Eq.(\ref{M_K_chi_K}) hold for the magnetic moment and susceptibility.

At low energies the "relativistic" dispersion law (\ref{eigenvalues}) emerges in a variety of physical systems \cite{Volovik}, and so the problem of the AB magnetism of a semimetal goes beyond graphene.  We will give a comprehensive treatment of the phenomenon by employing the zeta function regularization method \cite{zeta} that finds wide applications in calculations of the Casimir effect.   The flexibility and generality of the technique will allow us not only to solve the above-mentioned "relativistic" versions of the problem but also, at no extra cost,  discuss systems having more general dispersion laws.  As a by-product we will also consider the Landau diamagnetism in semimetals.    

Direct inspection of Eqs.(\ref{E_K}) and (\ref{E_K_ring}) shows that they are divergent.  The divergences are fictitious because the  expression for the spectrum (\ref{eigenvalues}) is only applicable at low energy;  furthermore, the sum and integral should only be over wavevectors within the first Brillouin zone.  AA \cite{Ando} treated this problem by introducing a cutoff function into Eq.(\ref{E_K}) which allowed them to carry out a numerical calculation of the magnetic moment.  From this they identified a cutoff-independent part which they argued captured the low energy part (\ref{eigenvalues}) of the true spectrum.  A compact derivation of AA's result will be given below as a special case of a more general theory.  

\section{Spectral zeta functions}

Our calculation follows the analysis of a similar problem \cite{Bogachek}.  We begin by defining the spectral zeta functions for the cylinder
\begin{equation}
\label{zeta_s_cyl}
\zeta_{M}^{\textrm{\textrm{(cyl)}}}(s)=\sum_{n=-\infty}^{\infty}\int_{-\infty}^{\infty}\frac{dq_{z}}{2\pi} \left [q_{z}^{2}+\left (\frac{2\pi}{W}\right )^{2}(n+\phi)^{2}+M^{2}\right ]^{-\frac{s}{2}}
\end{equation} 
and the ring
\begin{equation}
\label{zeta_s_ring}
\zeta_{M}^{\textrm{\textrm{(ring)}}}(s)=\sum_{n=-\infty}^{\infty} \left [\left (\frac{2\pi}{W}\right )^{2}(n+\phi)^{2}+M^{2}\right ]^{-\frac{s}{2}}
\end{equation}
versions of the problem.  Here $M$ confers a gap to the fermion spectrum, which will be put to zero later, and $s$ is a parameter.  For $M$ finite and $s$ positive and sufficiently large the expressions (\ref{zeta_s_cyl}) and (\ref{zeta_s_ring}) are convergent and can be explicitly evaluated.  The outcome will be analytically continued to the physically relevant situation of $M=0$ and $s=-1$.  This procedure extracts a cutoff-independent AB piece of the energies (\ref{E_K}) and (\ref{E_K_ring}) via $\mathcal{E}_{K}^{\textrm{\textrm{(cyl)}}}(\phi)=-\gamma L \zeta_{0}^{\textrm{\textrm{(cyl)}}}(-1)$ and $\mathcal{E}_{K}^{\textrm{\textrm{(ring)}}}(\phi)=-\gamma \zeta_{0}^{\textrm{\textrm{(ring)}}}(-1)$ which is all we need to compute magnetic properties (\ref{M_K_chi_K}).   The contribution that is dropped represents the cutoff-dependent ground-state energy of the filled Dirac sea in the absence of the AB flux; its value can be obtained by replacing the summation in Eqs.(\ref{E_K}) and (\ref{E_K_ring}) by an integration.

The spectral zeta functions (\ref{zeta_s_cyl}) and (\ref{zeta_s_ring}) contain the solutions to other problems, too.  Over fifty years ago Abrikosov and Beneslavskii \cite{AB} demonstrated that (in three dimensions) crystal symmetry permits both linear $s=-1$ (like in graphene)  and parabolic ("non-relativistic") $s=-2$ touching of the valence and conduction bands.  The latter parallels parabolic dispersion law found in unbiased bilayer graphene \cite{bilayer} if the interactions are neglected;  more generally, $s=-\nu$ describes a rhombohedral multilayer composed of $\nu$ graphene monolayers \cite{Volovik,RMP}.  Abrikosov and Beneslavskii additionally investigated the role of Coulomb interactions whose effect, like in graphene  \cite{Elias}, was shown to be fairly weak in the case of a linear spectrum.  However, Coulomb interactions have a dramatic consequence for the case of a parabolic spectrum, $s=-2$,  where a breakdown of single-particle description was predicted \cite{AB}.  The situation in bilayer graphene is similar where recent experimental and theoretical work \cite{experiment+theory} found that interactions can lead to a reconstruction of the ground state.  Additionally, the $s=-2$ case warrants special attention, because (i) the parabolic dispersion plays an important role in an explanation of the unconventional quantum Hall effect \cite{bilayer} and universal conductivity \cite{Katsnelson} in bilayer graphene, and (ii) it separates the regimes where the density of states is non-singular (for $-s <2$) vs. singular (for $-s>2$).  

The spectral zeta functions (\ref{zeta_s_cyl}) and (\ref{zeta_s_ring}) can be calculated by using the identity  \cite{integral}
\begin{equation}
\label{integral}
\int_{0}^{\infty}\frac{\cos px dx}{(x^{2}+a^{2})^{\frac{s}{2}}}=\sqrt{\pi}\left (\frac{p}{2a}\right )^{\frac{s-1}{2}}\frac{K_{\frac{s-1}{2}}(pa)}{\Gamma(\frac{s}{2})}, ~~\Re s>0
\end{equation}
where $\Gamma(z)$ is the Gamma function and $K_{\mu}(z)$ is the (MacDonald) modified Bessel function.  For $\Re s>1$ this permits integration of (\ref{zeta_s_cyl}) over $q_{z}$, leading to 
\begin{equation}
\label{integrated_over_q_z}
\zeta_{M}^{\textrm{\textrm{(cyl)}}}(s)=\frac{\Gamma(\frac{s-1}{2})}{2\sqrt{\pi}\Gamma(\frac{s}{2})}\zeta_{M}^{\textrm{\textrm{(ring)}}}(s-1)
\end{equation}
which relates the cylinder (\ref{zeta_s_cyl}) and ring (\ref{zeta_s_ring}) spectral zeta functions.  Thus knowledge of the full $s$ dependence of one of them describes both cases \cite{Nesterenko}.  For example, the AB magnetism of a graphene ring (described by $\zeta_{M=0}^{\textrm{\textrm{(ring)}}}(-1)$) can be inferred via (\ref{integrated_over_q_z}) from the zeta function for the cylinder (as $\zeta_{M=0}^{\textrm{\textrm{(cyl)}}}(0)$). 

When $\Re s>1$, the spectral zeta function for the ring (\ref{zeta_s_ring}) can be computed by employing the Poisson summation formula.  With the aid of (\ref{integral}) we find
\begin{eqnarray}
\label{zeta_ring_after_Poisson}
\zeta_{M}^{\textrm{\textrm{(ring)}}}(s)&=&\frac{4\sqrt{\pi}}{\Gamma(\frac{s}{2})}\left (\frac{W}{2\pi}\right )^{s}\nonumber\\
&\times&\Big\{\frac{\Gamma(\frac{s-1}{2})}{4}\left (\frac{MW}{2\pi}\right )^{1-s}+\left (\frac{2\pi^{2}}{MW}\right )^{\frac{s-1}{2}} \nonumber\\
&\times&\sum_{n=1}^{\infty}\frac{\cos 2\pi n\phi}{n^{\frac{1-s}{2}}}K_{\frac{s-1}{2}}(nMW) \Big\}
\end{eqnarray}
Combined with (\ref{integrated_over_q_z}) this provides us with an expression for the cylinder spectral zeta function  
\begin{eqnarray}
\label{zeta_cyl_after_Poisson}
\zeta_{M}^{\textrm{\textrm{(cyl)}}}(s)&=&\frac{2}{\Gamma(\frac{s}{2})}\left (\frac{W}{2\pi}\right )^{s-1}\nonumber\\
&\times&\Big\{\frac{\Gamma(\frac{s-2}{2})}{4}\left (\frac{MW}{2\pi}\right )^{2-s}+\left (\frac{2\pi^{2}}{MW}\right )^{\frac{s-2}{2}} \nonumber\\
&\times&\sum_{n=1}^{\infty}\frac{\cos 2\pi n\phi}{n^{\frac{2-s}{2}}}K_{\frac{s-2}{2}}(nMW) \Big\}
\end{eqnarray}
valid for $\Re s>2$.  These expressions can be analytically continued into the $\Re s<1$ and $\Re s<2$ regions, respectively, and the $M\rightarrow0$ limit can be taken which leads to our main results
\begin{equation}
\label{zeta_cylinder_main}
\zeta_{0}^{\textrm{\textrm{(cyl)}}}(s)=\frac{\Gamma(1-\frac{s}{2})}{\pi\Gamma(\frac{s}{2})}\left (\frac{2}{W}\right )^{1-s}\sum_{n=1}^{\infty}\frac{\cos2\pi n\phi}{n^{2-s}}
\end{equation}
\begin{equation}
\label{zeta_ring_main}
\zeta_{0}^{\textrm{\textrm{(ring)}}}(s)=\frac{2\Gamma(\frac{1-s}{2})}{\sqrt{\pi}\Gamma(\frac{s}{2})}\left (\frac{2}{W}\right )^{-s}\sum_{n=1}^{\infty}\frac{\cos 2\pi n\phi}{n^{1-s}}
\end{equation}

\subsection{Application: Aharonov-Bohm diamagnetism}

\subsubsection{Linear dispersion law: cylinder}

As a first application of Eq.(\ref{zeta_cylinder_main}) we consider the $s=-1$ case, which describes a cylinder with a linear dispersion law (\ref{eigenvalues}).  The part of the ground-state energy that depends on the AB flux will be given by 
\begin{equation}
\label{AB_energy}
\mathcal{E}_{K}^{\textrm{\textrm{(cyl)}}}(\phi)=-\gamma L\zeta_{0}^{\textrm{\textrm{(cyl)}}}(-1)=\frac{\gamma L}{\pi W^{2}}\sum_{n=1}^{\infty}\frac{\cos 2\pi n\phi}{n^{3}}
\end{equation}
The AB flux controls both the magnitude and sign of the result; the energy has maxima at $\phi$ integer and minima at $\phi$ half-odd integer.  The magnetic moment and susceptibility of the cylinder follow from Eqs.(\ref{M_K_chi_K}) as:
\begin{equation}
\label{M_K}
\mathcal{M}_{K}^{\textrm{\textrm{(cyl)}}}(\phi)=\frac{\gamma L}{2\pi \Phi_{0}}\sum_{n=1}^{\infty}\frac{\sin 2\pi n\phi}{n^{2}}\rightarrow -\frac{\gamma L}{\Phi_{0}}\int_{0}^{\phi}dt\ln2|\sin \pi t|
\end{equation}
\begin{equation}
\label{chi_K}
\chi_{K}^{\textrm{\textrm{(cyl)}}}(\phi)=\frac{\gamma W^{2}L}{4\pi \Phi_{0}^{2}}\sum_{n=1}^{\infty}\frac{\cos 2\pi n\phi}{n}\rightarrow -\frac{\gamma W^{2}L}{4\pi \Phi_{0}^{2}}\ln2|\sin \pi \phi|
\end{equation}
where the last two representations, valid in the $0<\phi<1$ range, should be periodically continued for all other $\phi$.  We observe that both quantities are proportional to the cylinder length $L$; the magnetic moment is independent of the circumference $W$, while the susceptibility is logarithmically divergent at $\phi$ integer.  This translates into a weak non-analyticity at integer values of $\phi$ for the energy (\ref{AB_energy}) and magnetic moment (\ref{M_K}), with the latter vanishing both at integer and half-odd integer $\phi$.  Eqs.(\ref{M_K}) and (\ref{chi_K}), combined with Eqs.(\ref{E_general}) for the total magnetic moment and susceptibility, reproduce the AA results \cite{Ando}.  

\subsubsection{Linear dispersion law: ring}

As an application of Eq.(\ref{zeta_ring_main}) we consider the $s=-1$ case which will describe a ring with linear dispersion law.  The AB piece of the ground-state energy will be given by
\begin{eqnarray}
\label{AB_energy_ring}
\mathcal{E}_{K}^{\textrm{\textrm{(ring)}}}(\phi)&=& -\gamma \zeta_{0}^{\textrm{\textrm{(ring)}}}(-1)=\frac{2\gamma}{\pi W}\sum_{n=1}^{\infty}\frac{\cos 2\pi n\phi}{n^{2}}\nonumber\\
&\rightarrow&\frac{2\pi\gamma}{W}\left (\frac{1}{6}-\phi+\phi^{2}\right )
\end{eqnarray}
where the last representation is valid for the range $0<\phi<1$ and should be periodically continued for all other $\phi$.  As in the case of the cylinder, the magnitude and sign can be controlled by the AB flux $\phi$.  The maxima of (\ref{AB_energy_ring}) are located at $\phi$ integer while the minima lie at half-odd integer $\phi$.  The magnetic moment then follows as
\begin{equation}
\label{M_K_ring}
\mathcal{M}_{K}^{\textrm{\textrm{(ring)}}}(\phi)=\frac{\gamma W}{2\Phi_{0}}(1-2\phi), ~~~0<\phi<1
\end{equation}
It is proportional to the circumference $W$,  varies between $\gamma W/2\Phi_{0}$ and $-\gamma W/2\Phi_{0}$, and has discontinuities at $\phi$ integer.   The susceptibility is constant and diamagnetic, except for positive delta-function peaks at $\phi$ integer.  

A remarkable feature of the ring geometry is that it displays a "ferromagnetic" broken symmetry at zero flux: the magnetic moment can be of either sign, depending on the history.  This was already implied by Eq.(\ref{E_K_ring}), which has discontinuous derivative at every integer value of $\phi$.  A possible experimental realization of such a ring represents an "armchair" graphene cylinder ($\alpha=0$ case of Eqs.(\ref{eigenvalues}) and (\ref{E_general})) whose length is much smaller than its circumference, $L\ll W$.  Such cylinders are not yet experimentally available but hopefully peculiarity of their ground state would stimulate efforts to produce them. 

\subsubsection{Parabolic and even power dispersion laws}

The case of the parabolic dispersion law $s=-2$ holds a surprise:  the cylinder and the ring spectral zeta functions (\ref{zeta_cylinder_main}) and (\ref{zeta_ring_main}) (as well as the corresponding magnetic moments) vanish at $s=-2$.  In fact, this remains true for any even $\nu=-s$ because this is where the Gamma function in the denominators of Eqs.(\ref{zeta_cylinder_main}) and (\ref{zeta_ring_main}) have poles. We thus conclude that the AB effect does not exist for a cylinder or ring with an even-layer rhombohedral graphene wall.  Inspection of Eqs.(\ref{zeta_ring_after_Poisson}) and (\ref{zeta_cyl_after_Poisson}) shows that for even $\nu=-s$ the AB effect is also identically zero in the presence of a gap which covers the case of a dielectric or a filled band of a metal ($s=-2$, $M\neq0$).  

\subsection{Application:  Landau diamagnetism}

The AB magnetism is physically closely related to the Landau diamagnetism as the latter is also due to currents circulating along the surface of the sample \cite{Kulik}.  The mathematical description of the two effects is also very similar.  Indeed, for a linear dispersion law the energy eigenvalues of the three-dimensional Dirac equation are given by \cite{Abrikosov} (compare with (\ref{eigenvalues}))
\begin{equation}
\label{Landau_spectrum}
E_{n}(q_{z})= \pm\gamma \left (q_{z}^{2} + \frac{2eH}{\hbar c}n\right )^{\frac{1}{2}}, n=0, 1, 2, ...
\end{equation}
where $q_{z}$ is parallel to the magnetic field $H$.  Each of the Landau levels (\ref{Landau_spectrum}) labeled by $n$ has a degeneracy $eHA/2\pi\hbar c$, where $A$ is the cross-sectional area of the sample perpendicular to the direction of the field.  In a semimetal, all negative energy Landau levels are filled while the $n=0$ state (shared between the valence and conduction bands) is half-filled.  With this in mind the ground-state energy can be written as (compare with (\ref{E_K}))
\begin{equation}
\label{E_3d}
\mathcal{E}^{(3d)}(H)=-\frac{\gamma eHA}{4\pi \hbar c}\sum_{n=-\infty}^{\infty}\int_{-\infty}^{\infty}\frac{Ldq_{z}}{2\pi}\left (q_{z}^{2} + \frac{2eH}{\hbar c}|n|\right )^{\frac{1}{2}}
\end{equation} 
where $L$ is the height of the sample in the direction of the magnetic field.  For a two-dimensional semimetal in a magnetic field perpendicular to the plane of the sample we would instead write (compare with (\ref{E_K_ring}))
\begin{equation}
\label{E_2d}
\mathcal{E}^{(2d)}(H)=-\frac{\gamma eHA}{4\pi \hbar c}\sum_{n=-\infty}^{\infty}\left (\frac{2eH}{\hbar c}|n|\right )^{\frac{1}{2}}
\end{equation}

\subsubsection{Spectral zeta functions}

As in the AB case, let us define the spectral zeta functions (compare with (\ref{zeta_s_cyl}) and (\ref{zeta_s_ring}) for $M=0$)
\begin{equation}
\label{zeta_3d}
\zeta_{0}^{(3d)}(s)=\sum_{n=-\infty}^{\infty}\int_{-\infty}^{\infty}\frac{dq_{z}}{2\pi} \left (q_{z}^{2}+\frac{2eH}{\hbar c}|n|\right )^{-\frac{s}{2}}
\end{equation}
\begin{equation}
\label{zeta_2d}
\zeta_{0}^{(2d)}(s)=\sum_{n=-\infty}^{\infty} \Big\{\left (\frac{2eH}{\hbar c}\right )^{2}n^{2}\Big\}^{-\frac{s}{4}}
\end{equation}
Comparing Eqs.(\ref{zeta_s_ring}) and (\ref{zeta_2d}) we notice that the latter can be analytically continued into the physically interesting region of negative $s$ by setting $\phi=0$ and replacing $2\pi/W\rightarrow2eH/\hbar c$ and $s\rightarrow s/2$ in Eq.(\ref{zeta_ring_main}):
\begin{equation}
\label{zeta_2d_main}
\zeta_{0}^{(2d)}(s)=\frac{2\Gamma(\frac{2-s}{4})}{\sqrt{\pi}\Gamma(\frac{s}{4})}\left (\frac{2eH}{\pi\hbar c}\right )^{-\frac{s}{2}}\zeta\left (\frac{2-s}{2}\right )
\end{equation}
where $\zeta(z)$ is the Riemann zeta function.  The part of the energy (\ref{E_2d}) per unit area dependent on the magnetic field is then given by
\begin{equation}
\label{E_H_2d}
\frac{\mathcal{E}^{(2d)}(H)}{A}=-\frac{\gamma eH}{4\pi\hbar c}\zeta_{0}^{(2d)}(-1)=\frac{\gamma \zeta(\frac{3}{2})}{16\pi^{2}}\left (\frac{2eH}{\hbar c}\right )^{\frac{3}{2}}
\end{equation} 
Multiplied by $4$ (graphene's degeneracy factor), this reproduces the result foreseen as early as 1989 \cite{1989}.

Since the spectral functions (\ref{zeta_3d}) and (\ref{zeta_2d}) satisfy the relationship (\ref{integrated_over_q_z}), the latter combined with (\ref{zeta_2d_main}) provides us with the analytic continuation of (\ref{zeta_3d}) into the region of physically interesting $s$:
\begin{equation}
\label{zeta_3d_main}
\zeta_{0}^{(3d)}(s)=\frac{\Gamma(\frac{s-1}{2})\Gamma(\frac{3-s}{4})}{\pi \Gamma(\frac{s}{2})\Gamma(\frac{s-1}{4})}\left (\frac{2eH}{\pi\hbar c}\right )^{-\frac{s-1}{2}}\zeta\left (\frac{3-s}{2}\right )
\end{equation}

\subsubsection{Bulk semimetal}

The physically relevant case now holds a surprise because at $s=-1$ the spectral zeta function (\ref{zeta_3d_main}) has a pole:
\begin{equation}
\label{pole}
\zeta_{0}^{(3d)}(s\rightarrow -1)\rightarrow-\frac{eH}{6\pi \hbar c}\frac{1}{s+1}\rightarrow-\frac{eH}{6\pi\hbar c}\ln\frac{L}{b}
\end{equation}
This is a sign of a logarithmic cutoff dependence;  the residue of the spectral function provides us with the amplitude of the logarithm \cite{KSLZ} as indicated in the last step.  Here $b$ is of the order of the interparticle spacing.  The magnetic piece of the energy (\ref{E_3d}) per unit volume is then given by
\begin{equation}
\label{E_H_3d}
\frac{\mathcal{E}^{(3d)}(H)}{AL}=-\frac{\gamma eH}{4\pi \hbar c}\zeta_{0}^{(3d)}(-1)=
\frac{v_{F}}{24\pi^{2}c}\frac{e^{2}}{\hbar c}H^{2}\ln\frac{L}{b}
\end{equation}  
where, to give a better idea of the magnitude of the effect, we substituted $\gamma=\hbar v_{F}$.  The magnetization $\mathcal{M}^{(3d)}$ and magnetic susceptibility $\chi^{(3d)}$ then follow as
\begin{equation}
\label{M_chi}
\mathcal{M}^{(3d)}(H)=\chi^{(3d)}H,~~~\chi^{(3d)}=-\frac{v_{F}}{12\pi^{2}c}\frac{e^{2}}{\hbar c}\ln\frac{L}{b}
\end{equation}
The last equation erroneously predicts that as $L\rightarrow \infty$, the susceptibility drops below the ideal diamagnetic limit of $-1/4\pi$.  This means that a more careful treatment is needed that distinguishes between the external magnetic field $H$ and the magnetic induction $B$ representing the field experienced by the electrons of the substance \cite{Abrikosov2}.  This can be accomplished by replacing $H$ with $B$ in Eq.(\ref{M_chi}) which no longer gives $\mathcal{M}(H)$;  the latter dependence can be found from the equation $H=B-4\pi\mathcal{M}(B)$.  As a result correct version of Eqs.(\ref{M_chi}) would read
\begin{equation}
\label{correct}
\mathcal{M}(H)=\chi H, ~~~~\chi=\frac{\chi^{(3d)}}{1-4\pi \chi^{(3d)}}
\end{equation}
where we assumed that a cylindrical sample is placed in an external axial magnetic field $H$.  We now see that in the thermodynamic limit $L\rightarrow \infty$ the susceptibility $\chi$ approaches $-1/4\pi$, i.e. the bulk semimetal is an ideal diamagnet.  In practice, however, we have $|\chi|\ll1$ and Eqs.(\ref{M_chi}) are adequate, as the amplitude of the logarithm in (\ref{M_chi}) is of the order $10^{-6}$; astronomically large sample sizes would be required to observe $|\chi|\approx1/4\pi$.

\section{Conclusions}

By employing zeta function regularization method we have given a comprehensive and unified description of the phenomena of the Aharonov-Bohm magnetism in semimetals in the ring and cylinder geometries for general dispersion laws and of the Landau diamagnetism in two- and three-dimensional semimetals.  While reproducing existing results as special cases of more general consideration, we have also arrived at a series of new predictions:

(i)  Nullification of the AB effect for "non-relativistic" and generally "even power" dispersion laws.

(ii) Ground-state ferromagnetic broken symmetry of a short "armchair" graphene tube at zero magnetic field.

(iii) Logarithmic sample size dependence of diamagnetic susceptibility of bulk semimetal.

\end{document}